\documentstyle[preprint,aps]{revtex}
\begin{document}
\draft
\title{Exact solution of the Sutherland model\\
with arbitrary internal symmetry}
\author{Yusuke Kato and Yoshio Kuramoto}
\address{Department of Physics, Tohoku University,\\
Sendai 980, Japan}
\maketitle
\begin{abstract}
An elementary theory is presented for solving the Sutherland model with
arbitrary internal symmetry such as SU($\nu$) or a supersymmetry SU($\nu,
\mu$).  The ground-state wave function and all the energy levels are derived.
One starts with solving a variant of the model with distinguishable particles,
and then (anti)symmetrizes the solution.  The theory is also applied to various
lattice versions of the model.  It is proved that the Gutzwiller-type wave
function is not only an eigenstate of the supersymmetric \it t-J \rm model, but
is indeed the ground state.
 \\
\end{abstract}
There has been an increasing interest in a family of one-dimensional models
with interaction proportional to inverse square of the distance.  With the
periodic boundary condition, the model is called the Sutherland
model\cite{Suth}.
Several years ago, Haldane\cite{Haldane1} and Shastry\cite{Shastry} found that
an $S=1/2$ spin chain with long-ranged exchange (HS model) is solvable and that
it is a lattice version of the original Sutherland model. After their works,
the charge degree of freedom, {\it hole}, was introduced into the HS model and
long-ranged supersymmetric {\it t-J} model has been proposed by Kuramoto and
Yokoyama\cite{KY}.
Ever after, the Sutherland model, HS model, and the {\it t-J} model were
generalized to multi-component systems and the generalized models have been
investigated intensively\cite{HaHaldane,Kawa1,MinahanPolychronakos}. In view of
fundamental physical interest in exact spin and charge dynamics in the {\it
t-J} model, for instance, it is highly desirable to develop a simple and
flexible theoretical scheme for systems with internal symmetries.

Sutherland and Shastry applied the asymptotic Bethe ansatz (ABA) to the
multi-component Sutherland model and derived the spectrum and degeneracy, and
discussed the thermodynamics\cite{S2}. In spite of its power and simplicity,
however, there are still unsolved issues in the ABA.
One is about the validity of the asymptotic region in the system with a finite
size. Another is about the ill-defined phase shift for equal momenta with
different spins.
Hence alternative approach is well worth a trial.

In this paper, we present an elementary theory for deriving all the energy
levels of the Sutherland model with arbitrary internal symmetry.
The main idea of our method is firstly to generalize the calculation in ref.
\cite{Suth} to a modified model for distinguishable particles. After that, we
take into account the internal degree of freedom of particles and symmetrize or
antisymmetrize the wave function to represent identical particles.  The present
theory has the following outstanding features. Firstly, our method identifies
easily the ground state and gives the explicit form of the wave function for
the most general models with the $SU(\nu,\mu)$ supersymmetry.
Secondly, it gives a microscopic derivation of the energy as a functional of
the momentum distribution function. Such an expression with correct account of
the degeneracy is important especially for investigation of thermodynamics.
Finally, it is also applicable to lattice models such as the multi-component
{\it t-J} model.  Because of these features, this theory should provide a basis
for further development such as investigation of dynamical properties for
systems with internal symmetries.

We consider the following model:
\begin{equation}
{\cal H}=
         -\sum_{i}\frac{\partial ^2}{\partial x_i^2}
         +\frac{2\pi^2}{L^2}\sum_{i<j}
         \frac{\lambda(\lambda -\zeta M_{ij})}{\sin^2
[\pi\left(x_{i}-x_{j}\right)/L]},\label{SP}
\end{equation}
where $M_{ij}$ is the exchange operator of coordinates of particles $i$ and $j$
\cite{Poly}. The size of the system is given by $L$, $\zeta$ is 1 or $-1$ and
the dimensionless coupling parameter $\lambda$ is positive. Since ${\cal H}$
does not depend on the internal symmetry of particles, we consider only the
orbital part of the wave function $\Psi \left(\left\{ x \right\}\right)$. For
the moment, we regard particles as distinguishable and does not impose any
permutation properties on the wave functions.
Here we write $\Psi$ in the form
$\Psi =\Psi_{0,\zeta}\Phi$
where $\Psi _{0,\zeta}$ is the ^^ ^^ absolute ground state", by which we mean
the ground state in the case where no restrictions to the symmetry of the
eigenfunction are imposed. For $\Psi _{0,\zeta}$ we can replace $\zeta M_{ij}$
in Eq.(\ref{SP}) by 1 since the wave function satisfying $\zeta M_{ij}\Psi
_{0,\zeta}=\Psi _{0,\zeta}$ minimizes the repulsion.  Then the result of
ref.\cite{Suth} for the single component system can be used, and explicit form
of $\Psi _{0,\zeta}$ is given by
\begin{equation}
\Psi _{0,\zeta}=\prod _{i<j}
                  \left|\sin \frac{\pi \left(x_i -x_j\right)}{L}
                    \right|
                     ^{\lambda}
                    \left(
                  {\rm  sgn}\left(x_i -x_j\right)
                  \right)^{\frac{1-\zeta}{2}},
\end{equation}
with the eigenenergy
\begin{equation}
E_{0}=\frac13\left(\frac{ \lambda \pi}{L}\right)^2
N\left(N^2-1\right).\label{abs}
\end{equation}
Here $N$ is the total number of particles.

Note that $\Psi_{0,\zeta}$ is the totally symmetric (antisymmetric) function
when $\zeta=1 \ (-1)$.
Considering $\Phi$ as a function of the variables
$z_{j}=\exp \left(i2\pi x_j/L\right)$,
we obtain the eigenvalue equation for $\Phi$:
\begin{equation}
{\cal H}'\Phi =\left[\sum_{j}\left(z_{j}\frac{\partial}{\partial
z_{j}}\right)^2+ \lambda \left(N-1\right)\sum_{j}z_j \frac{\partial}{\partial
z_{j}}+2 \lambda \sum_{j<k}h_{jk}\right]\Phi
              =\epsilon \Phi,
\end{equation}
with
\begin{equation}
h_{jk}=\frac{z_{j}z_{k}}{z_{j}-z_{k}}
\left[
  \left(
       \frac{\partial}{\partial z_j}-
       \frac{\partial}{\partial z_k}
  \right)
      -\frac{1}{z_j -z_k}\left(1-M_{jk}\right)
\right].
\end{equation}
The eigenvalue $\epsilon$ is related with the original eigenenergy $E$ by
$\epsilon =[L/(2\pi)]^2\left(E-E_{0}\right).$

Since the system has the translational symmetry, we consider eigenfunctions
with the total momentum $Q$.
We take a complete basis set of plane waves
\begin{equation}
\phi _{\kappa,P }=\prod _{i}z_{p\left(i\right)}^{\kappa _{i}},\quad\quad
\sum _{i}^{N}\kappa _{i}=Q,
\end{equation}
where $P=\left(p\left(1\right),p\left(2\right), \cdots ,p\left(N\right)\right)$
 is an $N$-th order permutation and $\kappa =\left(\kappa _{1}, \kappa _{2},
\cdots, \kappa _{N}\right)$ is a set of integers ordered so that $\kappa_{1}\ge
\kappa_{2}\ge \cdots \ge \kappa_{N} $.
The action of ${\cal H}'$ to $\phi _{\kappa ,P}$ gives
\begin{equation}
{\cal H}'\phi _{\kappa, P}=\sum _{i}\left[\kappa _{i}^2 + \lambda
\left(N-1\right)\kappa _{i}\right] \phi _{\kappa ,P}
+2\lambda
\sum_{j<k}h_{p\left(j\right)p\left(k\right)}\phi_{\kappa, P}.
\label{hprime}
\end{equation}
After some calculation, we obtain
\begin{equation}
h_{p(j)p(k)}\phi_{\kappa ,P}=-\kappa _{k}\phi_{\kappa
,P}+R\left(\left\{z_{p(i)}\right\}\right)\label{h},
\end{equation}
where $R$ is given by
\begin{equation}
R=\left\{
\begin{array}{cc}
\sum_{l=1}^{\kappa_{j}-\kappa_{k}-1}\left(\kappa_{j}-\kappa_{k}-l\right)\left(z_{p(k)}z_{p(j)}^{-1}\right)^{l}\phi_{\kappa, P}&\quad \left(\mbox{if $\kappa _{j}\ge \kappa _{k}+2$}\right),\\
0&\quad \left(\mbox{otherwise}\right).
\end{array}\right.
\label{R}
\end{equation}

We then define the order of the basis.
Let $\kappa '=\left(\kappa _{1}', \kappa _{2}', \cdots, \kappa _{N}' \right)$
be another set of momenta. We write $\kappa '<\kappa $ if the first
nonvanishing difference $\kappa _{i} -\kappa _{i}'$ is positive. For instance,
$(1,1,1,1,1)  < (2,1,1,1,0) < (2,2,1,0,0)$. Similarly we define the order of
the permutation $P=\left(p(1), p(2), \cdots ,p(n)\right)$. If $P'$ is another
permutation, we write $P'<P$ if the first nonvanishing difference $p(i)-p'(i)$
is positive. Each function of the basis is characterized by $\kappa $ and the
permutation $P$. We write $\left(\kappa ',P'\right) <\left(\kappa ,P\right)$ if
$\kappa '<\kappa $, or $\kappa '=\kappa $ and $P' <P$.
The off-diagonal elements of ${\cal H}'$ come from the second term of
Eq.(\ref{hprime}). From Eq.(\ref{R}) we see that $\langle \kappa',P'\vert {\cal
H}'\vert \kappa,P\rangle$ is zero when $\left(\kappa',P'\right) >
\left(\kappa,P\right)$.
Writing out ${\cal H'}$ in the ordered basis, the matrix is an upper triangular
one in which all matrix elements below the diagonal vanish. We can obtain all
the eigenenergies from the diagonal elements which come from the first terms of
the right hand side of Eqs.(\ref{hprime}) and (\ref{h}) as follows:
\begin{eqnarray}
\epsilon _{\kappa, P}
=\epsilon _{\kappa}
&=&\sum_{j}\left(\kappa_{j} ^2 + \lambda \left(N-1\right)\kappa_{j}\right)-2
\lambda \sum_{j<k}\kappa_{k}\nonumber\\
&=&\sum_{j}\left[\kappa ^2 _{j}+
\lambda\left(N+1-2j\right)\kappa_{j}\right].\label{epsilon}
\end{eqnarray}
Here we introduce the momentum distribution function $\nu \left( k \right)$ by
$\nu \left(k\right)=\sum_{j}\delta \left(\kappa_{j}, k\right)$
with $\delta \left(\kappa_{j},k\right)$ the Kronecker's delta symbol.
Then we can rewrite the expression (\ref{epsilon}) as a functional of $\nu
\left(k\right)$:
\begin{equation}
\epsilon _{\kappa}=\sum_{k=-\infty}^{\infty}k^2 \nu \left(k\right)
                  +\frac{\lambda}{2}\sum_{k,k'}\vert k-k'\vert
		   \nu \left( k \right)
		   \nu \left( k'\right)\label{functional}.
\end{equation}
All the eigenfunctions are written in the form:
\begin{equation}
\Phi_{\kappa ,P}=\phi _{\kappa, P}+\sum_{\kappa '< \kappa}\sum _{P'}a_{\kappa
',P'}\phi_{\kappa ',P'}\ .
\end{equation}
Each eigenvalue depends only on $\kappa $ and is independent of permutation
$P$. Thus, there is one-to-one correspondence between each eigenstate
$\Phi_{\kappa, P}$ of the present model and that of the free system
$(\lambda=0)$ given by $\phi_{\kappa,P}$.

Now we consider particles as $SU(\nu)$ fermions and calculate the ground state
under the given color distribution $\left\{M_{\sigma}\right\}_{\sigma=1}^{\nu}$
where $M_{\sigma}$ is the total number of particles with the color $\sigma$.
Firstly we consider the case $\zeta=1$. Since the absolute ground state $\Psi
_{0,1}$ is totally symmetric, the momenta of particles with the same color are
ordered non-equally $\left(\cdots\kappa_{j}>\kappa_{j+1}\cdots\right)$.
We introduce the momentum distribution of particles with the color $\sigma$ by
\begin{equation}
\nu _{\sigma}\left(k\right)=\sum_{j}\delta (\kappa_{j}, k)\delta(\sigma
,\sigma_{j}).
\end{equation}
In this case $\nu _{\sigma}\left(k\right)$ is either $0$ or $1$.
The energy relative to $E_{0}$ is given by Eq.(\ref{functional}) with $\nu
\left(k\right)=\sum_{\sigma}\nu _{\sigma}\left(k\right)$.
Let us consider the following momentum distribution:
\begin{equation}
\nu _{\sigma}\left( k\right)=
\theta \left(M_{\sigma}/2-\vert k\vert \right),
\label{distribution}
\end{equation}
where $\theta \left(k\right)$ is the step function,
and each $M_{\sigma}$ is taken to be odd.
It is clear that Eq.(\ref{distribution}) gives the minimum for both the first
and second terms in Eq.(\ref{functional}) as $SU(\nu)$ fermions.  Hence this
distribution gives the ground state.

Let $ \kappa _{g}$ be the set of the ordered momenta corresponding to the
momentum distribution (\ref{distribution}).
The ground-state wave function $\Phi _{g}$ is given by antisymmetrization of
$\Phi _{\kappa _{g},P}P\left(\left\{M_{\sigma}\right\}\right)\chi
\left(\left\{\sigma_{j}\right\}\right)$ where $\chi$ is a color function and
$P\left(\left\{\sigma_{j}\right\}\right)$ is given by%
\begin{equation}
P\left(\left\{\sigma_{j}\right\}\right)
=\prod_{\sigma=1}^{\nu}\delta\left(
			           M_{\sigma},\sum_{j=1}^{N}
                                   \delta\left(
				               \sigma_{j},\sigma
                                         \right)
                             \right).
\end{equation}
Firstly we antisymmetrize the coordinate of particles with the same color.
Because of the compact momentum distribution, the only part in $\Phi _{\kappa
_{g},P}$ that survives antisymmetrization is $\phi _{\kappa _{g},P}$.
 Using the formula for the Vandermonde determinant, the antisymmetrization
gives
\begin{equation}
\sum _{P}\left(-1\right)^{P}
\prod _{j=1}^{M_{\sigma}}z_{P(j)}^{-\left(M_{\sigma}+1\right)/2+j}
=\prod_{j=1}^{M_{\sigma}}z_{j}^{-\left(M_{\sigma}-1\right)/2}
 \prod_{j<k}^{M_{\sigma}}\left(z_{j}-z_{k}\right),
\end{equation}
where $\left(-1\right)^{P}$ denotes the sign of the permutation.  We must
multiply this by the antisymmetric color function for different colors.  The
result is given by
\begin{equation}
\Phi_{g}=\prod_{j=1}^{N}z_{j}^{-\left(M_{\sigma_{j}}-1\right)/2}\prod_{j<k}\left(z_{j}-z_{k}\right)^{\delta \left(\sigma_{j},\sigma_{k}\right)}P\left(\left\{M_{\sigma}\right\}\right)\exp \left[i\frac{\pi}{2} {\rm sgn} \left(\sigma_{j}-\sigma_{k}\right)\right].\label{groundstate}
\end{equation}
Ha and Haldane proved that this wave function is an eigenfunction of a model
equivalent to Eq.(\ref{SP}) with $\zeta =1$, and conjectured that it provides
the ground-state wave function\cite{HaHaldane}.
We have proved here that their conjecture is correct. To see the equivalence
between the models we note that $-M_{ij}$ in Eq.(\ref{SP}) can be replaced by
the color exchange operator $P_{ij}$ for $SU\left(\nu\right)$ fermions.
On the other hand if $M_\sigma$ is even the ground state is degenerate because
the highest momentum occupied can be either with plus or minus sign.  Provided
that $N/\nu$ is an odd integer,
the minimum energy distribution among all the color distribution
$\left\{M_{\sigma}\right\}$ is given by $M_{\sigma}=N/\nu$ for all $\sigma$'s.

Next we consider the case $\zeta=-1$ for $SU\left(\nu\right)$ fermions. In this
case the absolute ground state $\Psi_{0,-1}$ is totally antisymmetric and the
momentum distribution is bosonic: $\nu _{\sigma}\left(k\right)=0,1,2\cdots$.
The ground state is given by $\nu_{\sigma}\left(k\right)=M_{\sigma}\delta
\left(k,0\right)$ for all $\sigma $'s and the ground-state energy is just
$E_{0}$.
The corresponding eigenfunction is
$\Psi=\Psi_{0,-1}\left(\left\{z_{j}\right\}\right)
P\left(\left\{\sigma_{j}\right\}\right)$.
Any distribution $\left\{M_{\sigma}\right\}$ leads to the same energy given by
Eq.(\ref{abs}).  Hence there is a degeneracy
$
\left(N+\nu-1\right)!/[\left(\nu-1\right)! N!]
$
in the ground state.

In the cases where particles are $SU(\nu)$ bosons or the mixture of bosons and
fermions, the model given by Eq.($\ref{SP}$) can be solved in a similar way.
Thus we consider the most general $SU(\nu,\mu)$ model{\it, i.e.}, particles
consisting of $SU(\mu)$ fermions and $SU(\nu)$ bosons. The wave function for
the ground state has not yet been reported for this model.
In this case, we can rewrite $M_{ij}$ as
\begin{equation}
M_{ij}=\tilde P_{ij}\equiv \left\{
\begin{array}{cc}
-P_{ij}&\quad \mbox{(if both $i$- and $j$-th particles are fermions),}\\
 P_{ij}&\quad \mbox{(otherwise)}.\\
\end{array}\right.
\end{equation}

We first consider the case $\zeta=1$.
The absolute ground state is given by
\begin{equation}
\Psi _{0,1}=
\prod_{j<k}^{N^{\rm B}}\vert \xi _{j}-\xi _{k}\vert^{\lambda}
\prod_{j<k}^{N^{\rm F}}\vert \omega _{j}-\omega_{k}\vert^{\lambda}
\prod_{j}^{N^{\rm B}}\prod_{k}^{N^{\rm F}}\vert \xi
_{j}-\omega_{k}\vert^{\lambda}\label{absolute}
\end{equation}
where $\xi_{j} \left(\omega_{j}\right)$ and $\sigma_{j}^{\rm
B}\left(\sigma_{j}^{\rm F}\right)$ are variables for the complex coordinate and
the color of $j$-th boson (fermion), respectively, and $N^{\rm B}$($N^{\rm F}$)
is the total number of bosons (fermions).

Let the momentum distribution for bosons (fermions) with the color $\alpha$
$(\beta )$ be given by
$\left\{\nu_{\alpha }^{\rm B} \left(k\right)\right\}
\quad
\left(\left\{\nu_{\beta }^{\rm F}\left(k\right)\right\}\right)$, and the color
distribution by $\left\{M_{\alpha }^{\rm B}\right\}_{\alpha=1}^{\nu}\quad
\left(\left\{M_{\beta}^{\rm F}\right\}_{\beta=1}^{\mu}\right)$
where each $M_{\beta}^{\rm F}$ is taken to be odd.
Since the absolute ground state (\ref{absolute}) is symmetric,
$\nu_{\alpha}^{\rm B} \left(k\right)$'s take all non-negative integers :
$0,1,2,\cdots$ and $\nu_{\beta}^{\rm F}\left(k\right)$'s do $0$ or 1. The
minimum energy distribution is given by
\begin{equation}
\nu^{\rm B}_{\alpha}\left(k\right)=M_{\alpha}^{\rm B}\delta\left(k,0\right),
\quad
\nu^{\rm F}_{\beta}\left(k\right)=
\theta \left(M_{\beta}^{\rm F}/2-\vert k\vert\right).
\end{equation}
The corresponding wave function $\Psi $ is given by $\Psi _{0,1}\Phi_g$ where
\begin{eqnarray}
\Phi_g&=&
\prod_j^{N^{\rm F}}\omega_j^{-K\left(\sigma_j^{\rm F}\right)}
\prod_{j<k}^{N^{\rm
F}}\left(\omega_{j}-\omega_{k}\right)^{\delta\left(\sigma_{j}^{\rm
F},\sigma_{k}^{\rm F}\right)}\nonumber \\
&\times& \exp \left[\frac{i\pi}{2}{\rm sgn} \left(\sigma_{j}^{\rm
F}-\sigma_{k}^{\rm F}\right)\right]
P\left(\left\{M_{\alpha}^{\rm B}\right\}\right)
P\left(\left\{M_{\beta}^{\rm F}\right\}\right),\label{suground}
\end{eqnarray}
with $K(\sigma) = (M_\sigma^{\rm F,B}-1)/2$.
Similarly, the ground state in the case $\zeta =-1$ is given by $\Psi
_{0,-1}\Phi _g '$ where
\begin{eqnarray}
\Phi_g '&=&
\prod_j^{N^{\rm B}}\xi_j^{-K\left(\sigma_j^{\rm B}\right)}
    \prod_{j<k}^{N^{\rm
B}}\left(\xi_{j}-\xi_{k}\right)^{\delta\left(\sigma_{j}^{\rm B},\sigma_{k}^{\rm
B}\right)}\nonumber\\
&\times&\exp \left[\frac{i\pi}{2}{\rm sgn} \left(\sigma_{j}^{\rm
B}-\sigma_{k}^{\rm B}\right)\right]
P\left(\left\{M_{\alpha}^{\rm B}\right\}\right)
P\left(\left\{M_{\beta}^{\rm F}\right\}\right).\label{suground2}
\end{eqnarray}

Our method is also applicable to lattice models by taking the limit $\lambda
\rightarrow \infty$ \cite{S2,Haltani}.
In the strong coupling limit, a part of the potential $\lambda^2\sum
_{j<k}{\sin ^{-2}\left[\pi \left(x_j -x_k\right)/L \right]}$ enforces particles
to localize with a lattice spacing $L/N$.
Up to order ${\cal O}\left(\lambda\right)$, the Hamiltonian decouples into
\begin{equation}
{\cal H}={\cal H}_{\rm ela}+{\cal H}_{\rm lat}+E_{\rm Mad},\label{infinit}
\end{equation}
where ${\cal H}_{\rm ela}$, ${\cal H}_{\rm lat}$, and $E_{\rm Mad}$ are the
elastic and lattice Hamiltonians and the Madelung energy, respectively. We
define $u_{j}$ as the displacement from the $j$-th lattice point $(j=1,2,\cdots
,N)$ and introduce a function
\begin{equation}
\tilde d \left(j\right)=3\sin ^{-4}\left(\frac{\pi j}{N}\right)
-2\sin ^{-2}\left(\frac{\pi j }{N}\right).
\end{equation}
Then we obtain
\begin{equation}
{\cal H}_{\rm ela}=-\sum _{j}\frac{\partial^2}{\partial u_{j}^2}
         +2\lambda^2\left(\frac{\pi}{L}\right)^4\sum_{j}\left[\sum_{k\left(\ne
j\right)}\tilde d\left(j-k\right)\right]u_{j}^2
         -4\lambda^2\left(\frac{\pi}{L}\right)^4\sum_{j<k}\tilde
d\left(j-k\right)u_{j}u_{k}, \label{elastic}
\end{equation}
where $j$ and $ k$ are no longer indices of particles but those of sites.
We rewrite ${\cal H}_{\rm ela}$ as
\begin{equation}
{\cal H}_{\rm ela}=\sum_{q=0}^{N-1}4\lambda
\left(\frac{\pi}{L}\right)^2q\left(N-q\right)\left(\hat b_{q}^{\dagger}\hat
b_{q}+\frac12\right),\label{elastic2}
\end{equation}
where $\hat b_{q}^{\dagger}\ \left(\hat b_{q}\right)$ is the creation
(annihilation) operator of the phonon field.
In deriving the expression (\ref{elastic2}), we have used the following result:
\begin{equation}
\sum_{l=1}^{N-1}\tilde d\left(l\right)
\exp \left(\frac{i2\pi ql}{N}\right)
=-2q^2\left(q-N\right)^2+\frac{N^4-1}{15}.
\end{equation}
The rest of ${\cal H}$ is given by
\begin{equation}
E_{\rm Mad}=2\left(\frac{\pi}{L}\right)^2\sum_{j<k}\frac{\lambda^{2}}{\sin
^{2}[\pi \left(j-k\right)/N]}
=\frac13\left(\frac{\pi \lambda}{L}\right)^{2}N\left(N^2-1\right),
\label{Mad}
\end{equation}

\begin{equation}
{\cal H}_{\rm lat}=-2\left(\frac{\pi}{L}\right)^2\sum_{j<k}\frac{\lambda \zeta
\tilde P_{jk}}{\sin ^{2}[\pi \left(j-k\right)/N]}, \label{lattice}
\end{equation}
where $\tilde P_{jk}$ is the color exchange operator between the $j$-th and
$k$-th sites.
When $\left(\zeta, \nu,\mu \right) = \left(1,0,2 \right)$ or
$\left(-1,2,0\right)$, ${\cal H}_{\rm lat}$ becomes the Hamiltonian of the HS
model. In the case of $\left(\zeta, \nu,\mu\right)=\left(1,1,2\right)$ or
$\left(-1, 2, 1\right)$, it becomes the Hamitonian of the supersymmetric {\it
t-J} model.

We now derive the wave function $\Psi _{g}=\Psi _{0,1}\Phi_g$ of the ground
state with $\zeta = 1$ in the limit of $\lambda \rightarrow \infty$. The
absolute ground state $\Psi _{0,1}$ becomes
\begin{equation}
\Psi_{0,1} \rightarrow
\exp \left[-\lambda
\left(\frac{\pi}{L}\right)^2\sum_{q=0}^{N-1}q\left(N-q\right)u\left(q\right)u\left(N-q\right)\right]\times \left(\mbox{numerical factor}\right),
\end{equation}
where $u\left(q\right)$ is the Fourier transform of the displacement $u_{j}$.
The $\Psi _{0,1}$ corresponds to the zero-point motion of the phonon field.
Factoring out the phonon part, we obtain the ground state of the $SU(\nu, \mu)$
lattice model, which is nothing but $\Phi_g$
given by Eq.(\ref{suground}).
The remaining effect of $\Psi _{0,1}$ is the exclusion of multiple occupation
of each site.
In the special case of $SU(1,2)$, it is therefore proved that the Gutzwiller
wave function is indeed the ground state of the {\it t-J} model \cite{KY}.

Thus the present theory gives a unified treatment of the family of
Sutherland-type models.
In every case, our theory derives all the energy levels by a simple
calculation.  We have proved in this way the previous conjectures on the ground
states.
Furthermore simplicity of the method permits us to derive new results for the
ground-state wave function of $SU(\nu,\mu)$ models in both the continuum space
and the lattice.

Lastly, we make a brief remark on another application of our theory.
Similar approach is useful to the Calogero model with the internal symmetry as
well \cite{Vacek}. As a matter of fact, the Calogero model is much more
tractable than the Sutherland model. For the Calogero model, the local
operators for energy boost were already found \cite{Poly,Brink}. By using such
operators, we have obtained all the eigenstates of the $SU(\nu )$ Calogero
model and proved that the eigenfunction proposed
in refs. \cite{MinahanPolychronakos,Vacek}
is the exact ground state. The details will be discussed elsewhere.

\end{document}